\documentclass[pra,aps,amsfonts,amsmath,superscriptaddress,%
  twocolumn,showpacs,floatfix]{revtex4}
 
\def\be{\begin{eqnarray}}   
\def\ee{\end{eqnarray}}   

\usepackage{amsfonts}
\usepackage{amsmath}
\usepackage{bm}
\usepackage{epsfig}
\usepackage{xcolor}

\begin{document}
\title{Constraining the nuclear pairing gap with pairing vibrations}

\author{E. Khan} 
\address{
 {\it Institut de Physique Nucl\'eaire, Universit\'e Paris-Sud, 
IN2P3-CNRS, F-91406 Orsay Cedex, France}
}

\author{M. Grasso} 
\address{
 {\it Institut de Physique Nucl\'eaire, Universit\'e Paris-Sud, 
IN2P3-CNRS, F-91406 Orsay Cedex, France}
}

\author{J. Margueron} 
\address{
 {\it Institut de Physique Nucl\'eaire, Universit\'e Paris-Sud, 
IN2P3-CNRS, F-91406 Orsay Cedex, France}
}

\begin{abstract}

Pairing interactions with various density dependencies (surface/volume
mixing) are constrained with the two-neutron separation energy in the Tin
isotopic chain. The response associated with pairing vibrations in very
neutron-rich nuclei is sensitive to the density dependence of the pairing
interaction. Using the same pairing interaction in nuclear matter and in Tin
nuclei, the range of densities where the LDA is valid in the pairing channel
is also studied.
\end{abstract}
\maketitle

\section{Introduction}

Studies on pairing effects in both nuclear matter and finite nuclei have
known intensified interests in the recent years \cite{dea03}. There are two
main approaches for pairing, depending whether the mean field is based on
Gogny finite range interaction or on Skyrme interaction. In the first
approach, a similar functional is used in both the particle-hole channel and
the pairing channel, although interactions are not exactly the same due to
the density dependence of the pairing interaction. In the Skyrme approach,
the functionals are meant to be different in the two channels, as witnessed
for instance by their density dependence. The use of a different interaction
in the particle-hole channel and in the pairing channel has been justified a
decade ago \cite{gar99}; this is for instance the case of employing the
Skyrme interaction in the particle-hole channel and a zero range density
dependent interaction in the pairing channel. We shall focus on the Skyrme
approach: in this case the pairing density functional is difficult to
constrain and it has not been possible to derive an universal pairing
interaction during past decades, using for instance the odd-even mass
staggering on finite nuclei (see e.g. \cite{doba,gor07}). This may indicate the
need for another approach, using additional constrains: should the pairing
density functional be extended, and are there additional relevant
observables to constrain it ?

Nuclear matter could help in constraining the pairing functional. This
requires however to bridge nuclei and nuclear matter through LDA in the
pairing channel: its condition of validity should be more systematically
analysed. It has been recently shown that the two paired neutrons are
spatially localised in low density medium which corresponds to the surface
of the nucleus \cite{pil07}. The same conclusion is drawn by analysing the
di-neutron configuration in the excited states \cite{mat05,hag07a}, and also
performing calculations in low density matter in \cite{mat06,hag07}, mainly
renewing the possibility to link in some cases the nuclear matter and nuclei
in the pairing channel through the LDA.

Concomitantly the pairing functional has been extended in order to study the
condensation of the Cooper pairs (BEC-BCS crossover) in both symmetric and
neutron matter. In nuclear matter the medium polarization increases the
pairing gap at low densities in symmetric matter, whereas it reduces the gap
in neutron matter, indicating an isospin dependence of the pairing
functional \cite{mar07}. The application to finite nuclei of extended
pairing density functional have shown the relevance of the LDA in the
pairing channel \cite{mar08}. 

The pairing functional studies may thus enter in a new era, renewing the
method to design the pairing interaction: i) using an isospin dependence of
the pairing interaction ii) using eventually the nuclear matter as an additional
constrain for the pairing interaction iii) looking for additional
observables in nuclei than the odd-even mass staggering to constrain the
pairing interaction. Point i) has been investigated in \cite{mar07,mar08}.
Point ii) requires the validity of the LDA in the pairing channel and
shall be addressed in the present work. 

In the case of point iii) an interesting observable is pairing vibrations,
measured through two-particle transfer. It is well known that the transfer
cross section crucially depends on the pairing interaction at work in the
transferred paired \cite{bro73,oer01}. However in the 70-80's the form
factor of the transition has never been calculated fully microscopically.
The first microscopic calculations has been performed only recently
\cite{kha04}, allowing for a strong link between the pairing interaction and
pairing vibrations. Several calculations followed \cite{mat05,ave08},
showing the renewed interest for such studies.

It is therefore meaningful to use pairing vibrations as a complementary
observable to the masses, in order to constrain the pairing interaction, and
study the implications to the nuclear matter. More generally, the isospin
extension of the pairing interaction should be accompanied with additional
constraints. One purpose of this work is to evaluate if pairing vibrations
could play this role (Section III).

The method is to analyse the sensitivity of pairing vibrations to various
pairing interactions which provide the same two-neutron separation energy in
Tin isotopes, and evaluate the consequences on the pairing gap in symmetric
and neutron matter. On this purpose it is necessary to determine the range
of density where the LDA is valid in the pairing channel (Section II).

\section{Validity of the LDA in the pairing channel}

After many years of study, there is still no unambiguous universal pairing
functional ranging on the whole nuclear chart, and current efforts are
aiming in that direction. The problem may be due to the method used to
constrain it, namely comparing the pairing gap with odd-even mass
differences, or evaluating the separation energies along a given isotopic
chain. It therefore may be useful to consider a more general context: the
evaluation of several pairing interactions constrained by odd-even mass
difference, on nuclear matter on one side, and on additional observables on
the other side, should shed a renewed light on the problem. To achieve this
goal it is first necessary to determine the range of validity of the LDA in
the pairing channel.

\subsection{Method to determine the functional}

The method is the following: we first consider surface and various mixed paring
interactions. The parameters are determined so as to describe the two
neutron separation energy. Then pairing vibrations are used in order to
disentangle between the various pairing interactions (Section III). We choose $^{124}$Sn
and $^{136}$Sn nuclei: these are spherical nuclei where pairing vibrations
are likely to occur \cite{oer01}. One is stable and the second has a large
neutron excess.

The microscopic calculations for the ground state are based on the 
Hartree-Fock-Bogoliubov (HFB) model. The Skyrme interaction SLy4
\cite{cha98} is chosen for the particle-hole channel of the HFB equations.
The adopted pairing interaction is 
the usual zero-range density-dependent interaction 
\begin{equation}\label{eq:vpair}
 V_{pair}=V_0\left[1-\eta\left(\frac{\rho(r)}{\rho_0}\right)^\alpha\right]
 \delta\left({\bf r_1}-{\bf r_2}\right)
\end{equation}

where $\eta$ provides the surface/volume character of the 
interaction. We set 
$\alpha=1$ and
$\rho_0=0.16$ fm$^{-3}$. The numerical cutoff for the microscopic 
calculations is given by 
$E_{max}=60$ MeV (in quasiparticle energies) and $j_{max}$= 15/2. 
For each value of $\eta$, $V_0$ is chosen to fit the 
known experimental two-neutron separation energies for Sn isotopes. 
Surface and mixed interactions have been considered in this work 
and the used values of 
($\eta$,$V_0$) are listed in Table I. 

\begin{table}[!t]
  \caption{\label{inte} Values of $\eta$ and $V_0$ of the pairing interaction.}
    \begin{ruledtabular}
      \begin{tabular}{ccccc}
  &        $\eta$         & $V_0$ (MeV fm$^{-1}$)    \\
       \hline
&	0.35 & -285 \\
     &   0.65 &  -390     \\
    &    1 &   -670     \\
      \end{tabular}
    \end{ruledtabular}
\end{table}

\begin{figure}
\begin{center}
\epsfig{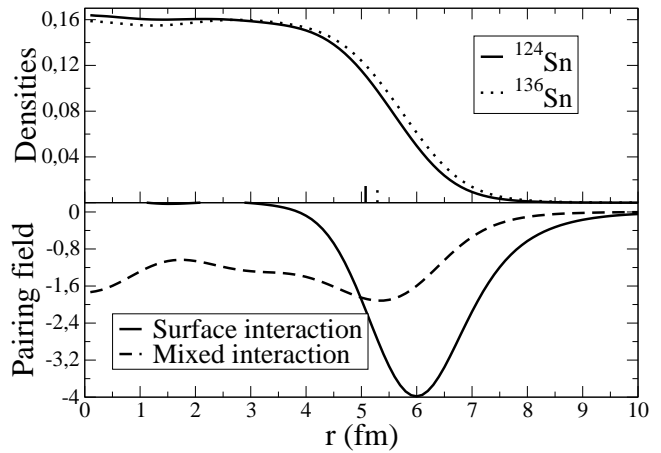}
\end{center}
\caption{{\it Upper panel:} Matter densities calculated with the HFB model for
$^{124}$Sn and $^{136}$Sn. The vertical lines indicate the radius corresponding the 
density at which all the pairing interactions converge in uniform matter (see
text). {\it Lower panel:} pairing field of $^{124}$Sn calculated with a surface $\eta=1$ 
(top) and a mixed $\eta=0.35$ (bottom) pairing interaction. }
\label{fig:densi}
\end{figure}

As an illustration to visualize the features of the calculated pairing
effects, we display in Fig. \ref{fig:densi} the neutron pairing field for
$^{124}$Sn corresponding to the surface $\eta=1$ and the mixed $\eta=0.35$
interactions.

\subsection{Pairing gap in uniform matter}

The relation between the pairing gap in uniform matter at a given density
and the pairing field at a given radius in nuclei has been explored in
Ref.~\cite{mar08}. It has been found that in the case of mixed interactions,
the LDA is in good agreement with the full microscopic HFB calculation
(differences less than 15\% on the pairing field). This might be related to the
extension of the Cooper pair which is getting smaller at the surface of
nuclei (about 2 fm) compared to that in the interior (about 5-6 fm)
\cite{pil07}. Close to the surface, pairing properties shall not be
very different from that of a uniform piece of matter at the same density.
It is then interesting to explore the low density properties of the
different pairing interactions listed in Table I. 

\begin{figure}
\begin{center}
\epsfig{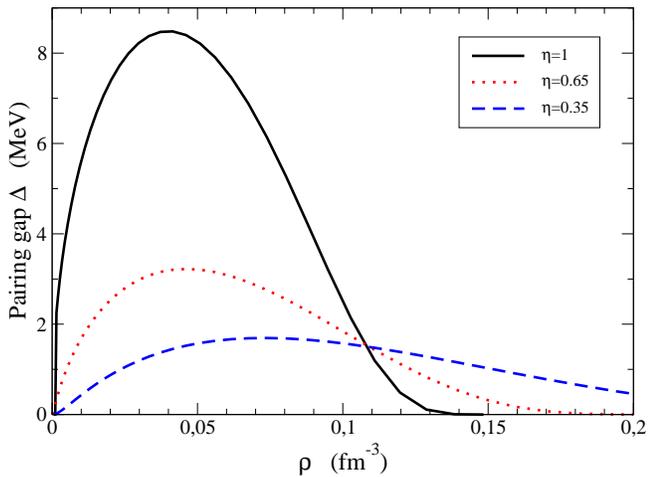}
\end{center}
\caption{Pairing gap versus the density for uniform matter
for different pairing interactions.}
\label{fig:unif}
\end{figure}

Fig.~\ref{fig:unif} displays the pairing gap in uniform matter for various
pairing interactions. It is observed that the different interactions leads
to very different pairing gap at low density while around saturation
density, there is a density ($\rho$=0.11~fm$^{-3}$) at which the pairing gap
and pairing strength coincide for the three pairing interactions. 

From Fig.~\ref{fig:unif}, two conclusions can be drawn: i) the two-neutron
separation energy used to adjust the parameters of the pairing interaction
is an observable which provides a strong constrain on the pairing gap
localized at the surface of the nuclei: the pairing gap in nuclear matter
has been constrained for $\rho$=0.11 fm$^{-3}$, which corresponds to R
$\simeq$ 5 fm in Tin nuclei, as showed by the $^{124}$Sn densities displayed
on Fig. ~\ref{fig:densi} ii) to better constrain the value of the parameter
$\eta$, one shall find another observable sensitive to the pairing strength
at low density (large radius, experimentally easier to probe).  Indeed, in
the very external part of the nuclei the pairing strength is very different
from one interaction to another. The pure surface pairing interaction
predict a pairing gap as high as 8~MeV at low density while the various
mixed pairing interaction are grouped below 3~MeV (see Fig.~\ref{fig:unif}).

Therefore, one might expect that properties of collective modes sensitive to
the external part of the nuclei could be changed by the properties of the
pairing interaction at low density. Pair transferred reaction mechanisms
like (p,t) or ($\alpha$, $^6$He) which is very surface peaked shall also help in
extracting the value of the pairing gap in the external part of the nuclei
or equivalently at low density.

\subsection{Validity of the LDA}

One aim of this work is to investigate the mapping of the pairing gap
between nuclei and nuclear matter, implying the following relation:

\begin{equation}\label{eq:ldapair}
\Delta_\infty(\rho)=\Delta_{HFB}(\rho(r)=\rho) 
\end{equation}

Eq. (\ref{eq:ldapair}) corresponds to the LDA in the pairing channel. However,
it should be noted that in practice, the LDA might still be valid even if
Eq. (\ref{eq:ldapair}) is not verified for very low densities ($\rho \le$ 0.1
fm$^{-3}$): the major role is played by densities around the saturation
density, where nucleus properties are the most important.

\begin{figure}
\begin{center}
\epsfig{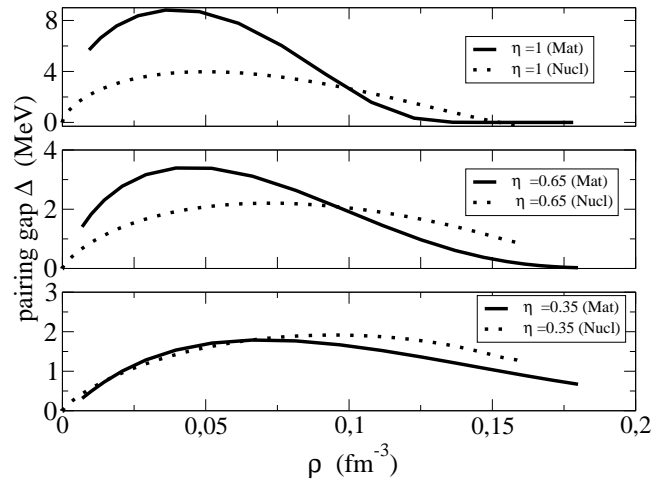}
\end{center}
\caption{Pairing gap versus the density for nuclear matter
and for $^{124}$Sn (see text)}
\label{fig:lda}
\end{figure}

Fig. ~\ref{fig:lda} displays the gap calculated with the HFB approach as a
function of the nuclei density in $^{124}$Sn. Nuclear matter gap is
surimposed, where the neutron-proton asymmetry are taken as the one of
$^{124}$Sn. One can see that the LDA is not valid for a pure surface
pairing, but becomes valid in the $\eta$=0.35 mixed case, for densities
lower than 0.1 fm$^{-3}$. Therefore, in this case, the low density part of
the nuclear matter gap could be constrained by nuclei calculations of the
gap: a valid case of the LDA in the pairing channel is showed here
quantitatively for the first time. This range of density corresponds to a
radius larger than about 5.1~fm in $^{124}$Sn and 5.2~fm in $^{136}$Sn
localised at the surface of the nuclei, as can be seen on
Fig.~\ref{fig:densi}. In the cases of the other pairing interactions, the
LDA underestimates the pairing gap at low density and overestimate it at
high density. However, it should be noted that the relative difference of the
gaps between the various pairing interactions are respected.

As noted above, to assess in practice the validity of the LDA, the density
range around the saturation density should be investigated, where nuclei
bulk properties are at work. Therefore a large discrepancy at very low
density ($\rho \le$ 0.1 fm$^{-3}$), as observed in the case of the surface
pairing interaction in Fig .~\ref{fig:lda}, does not preclude on the
validity of the LDA. More generally it can be stated that the less important
the surface contribution to the pairing interaction, the better the LDA.

\section{Pairing vibrations}

As stated above, it may be useful to consider an additional observable than
the separation energy, in order to constrain the pairing interaction, namely
its density dependence. There are only few observables which could be
relevant to constrain pairing effects. It has been shown that the first
2$^+$ state in nuclei is sensitive to the pairing interaction \cite{kha02}:
both its position and strength depend of the pairing interaction. However
this is mainly related to the pairing gap value, which is the same
observable extracted from odd-even mass excess. It should be noted than none
of these two observables (the first 2$^+$ state and the odd-even mass
staggering) can be directly linked to predictions. On one side there is the
difficulty to modelize excited states. On the other side, the difficulty is
to describe odd nuclei.

Pairing vibrations may be more adequate observable. They can be probed for
instance with two neutron transfer in nuclei close to shell closure. We
refer to \cite{oer01,bro73} for details on pairing vibrations.
Basically, these modes corresponds to the (collective) filling of subshells,
in transition from an A to A+2 nuclei. 

With pairing vibrations, pairing effects are probed by 3 ways. The first one
is the magnitude of the pairing gap $\Delta$ (average of the pairing field): a
large pairing gap implies strength at larger energies, following the formula
E$^2\simeq$($\epsilon - \lambda$)$^2$+ $\Delta^2$. This component is also present
in the first 2$^+$ state in the ph response as well as in the odd-even mass
staggering. But in the case of the pairing vibrations, there are two
additional contributions: first, the transition densities generating the
strength are the pairing one, which means that the unperturbed response as
well as the perturbed response are sensitive to the impact of the pairing
on the wave functions. Finally, the residual interaction, generating the
QRPA response, is also sensitive to pairing.

Pairing vibrations are therefore expected to be very sensitive to the
pairing interaction. On the other hand, it may also be difficult to
disentangle between the three above mentioned effects. However, the first
one can be evaluated using the energies of the unperturbed response, the
second one by studying the pairing transition densities, and the last one by
comparing the unperturbed and the QRPA responses. It should be noted that
a related study will also be performed in \cite{mat09b}.

\subsection{Method: QRPA in the pp channel}

The method is described in \cite{kha02,kha04}. Namely the QRPA equations are
solved in coordinate space, using the green function formalism. The
variation of the generalized density ${\cal R}$' is expressed in term of 3
quantities, namely $\rho'$, $\kappa'$ and $\bar{\kappa}'$, which are written
as a column vector: 

\begin{equation}\label{eq:rhodef}
          \rho' =  \left(
	\begin{array}{c}
	\rho' \\
         \kappa' \\
	 \bar{\kappa}' \\
        \end{array}
	\right),
	\end{equation}

where $\rho'_{ij} = \left<0|c^{\dagger}_jc_i|'\right>$
is the variation of the particle density,
$\kappa'_{ij} =\left<0|c_jc_i|'\right>$ and $\bar{\kappa}'_{ij} =
\left<0|c^{\dagger}_jc^{\dagger}_i|'\right>$ are the
fluctuations of the pairing tensor associated to the pairing vibrations and
$\mid ' \rangle$ denotes the change of the ground state wavefunction $|0>$
due to the external field. In contrast with the RPA where one needs
to know only the change of the ph density ($\rho'$), the variation of the
three quantities (\ref{eq:rhodef}) have to be calculated in the QRPA. In the
three dimensional space introduced in Eq. (\ref{eq:rhodef}), the first
dimension represents the particle-hole (ph) subspace, the second the
particle-particle (pp) one, and the third the hole-hole (hh) one. The
response matrix has therefore 9 coupled elements in QRPA, compared to one in
the RPA formalism.

The variation of the HFB Hamiltonian is given by:

\begin{equation}\label{eq:hvar}
 H'=	\bf{V}\rho',
\end{equation}
where $\bf{V}$ is the matrix of the residual interaction 
expressed in terms of the second derivatives of the HFB energy 
functional, namely:
					   
\begin{equation}\label{eq:vres}
{\bf{V}}^{\alpha\beta}({\bf r}\sigma,{\bf r}'{\sigma}')=
\frac{\partial^2{\cal E}}{\partial{\bf{\rho}}_\beta({\bf r}'{\sigma}')
\partial{\bf{\rho}}_{\bar{\alpha}}({\bf r}\sigma)},~~~\alpha,\beta = 1,2,3.
\end{equation}
In the above equation the notation $\bar{\alpha}$ means that whenever $\alpha$ is 2 or 3
then $\bar{\alpha}$ is 3 or 2.

The QRPA Green's function G can be used for calculating the strength function
associated with the two-particle transfer from
the ground state of a nucleus with A nucleons to the excited states of a 
nucleus with A+2 nucleons. This strength function is :

\begin{equation}\label{eq:stren2}
S(\omega)=-\frac{1}{\pi}Im \int  F^{*}({\bf r}){\bf{
G}}^{22}({\bf r},{\bf r}';\omega) F({\bf r}')
d{\bf r}~d{\bf r}'
\end{equation}
where ${\bf{G}}$ denotes the (pp,pp) component of the 
Green's function and F in the external perturbating field associated with the
addition of two particles.

In the QRPA calculations the full HFB quasiparticle spectrum up to 60 MeV is
included. These states are used to construct the unperturbed Green's
function ${\bf G_{0}}$. The residual interaction is derived from the
two-body force used in HFB according to Eq. (\ref{eq:vres}). The
contribution given by the velocity-dependent terms of the Skyrme force to
the residual interaction is calculated in the Landau-Migdal approximation,
which is shown to be accurate \cite{mat09}. The strength function for the
two-neutron transfer is calculated using Eq. (\ref{eq:stren2}). The
unperturbed Green's function is calculated with an averaging interval equal
to 0.15 MeV. All details can be found in Ref. \cite{kha04}.

The response function is calculated for the pp channel. All the calculations
are performed in a box of size 22.5 fm. It should be noted that exact
continuum treatment is much heavier, especially for nuclei such as Sn
isotopes. Moreover the aim is not to study the impact of the continuum
treatment (see \cite{kha04} for such a study). Finally the Sn isotopes under
study are far from the drip line, and continuum effects are expected to play
a negligible role.

\subsection{Unperturbed response results}

The HFB solutions are used in the quasiparticle random-phase approximation 
(QRPA) scheme to analyze self-consistently the excitations modes associated
to the pair transfer reactions. Since we study here two-neutron transfers, 
we focus on the neutron HFB quasiparticle states that are used to construct 
the elementary configurations of the excited modes.  We work with
positive-energy quasiparticle states.  Once calculated the quasiparticle
spectrum, it is possible to deduce some properties of the unperturbed
response function.

The quasiparticle states with energy less than 6 MeV and an occupation
probability $\le$ 80 \% are presented in Tables II and III for $^{124}$Sn
and $^{136}$Sn, respectively. Let us discuss the two cases $\eta=0.35$ and
$\eta=1$ (for $\eta=0.65$, results are similar to those obtained with
$\eta=0.35$).  For $^{124}$Sn, in the case of a mixed pairing interaction,
$\eta=0.35$, all the quasiparticle states with energy lower than 5 MeV are
totally occupied with the exception of a $h_{11/2}$ state at 1.5 MeV which
is 42\% occupied. This is the only low-energy state that can contribute to
some extent to the excitation mode.  The states that are completely empty
and can thus contribute more to the excitation are located at higher
energies. The first is an $f_{7/2}$ state at 5.8 MeV. The others have larger
energies (at least 1 MeV more).  One can thus expect that the unperturbed
response profile starts with a peak at twice 5.8 MeV, i.e. at $\sim$ 11.6
MeV (with some small contribution at 3 MeV).  In the case of a surface
interaction, $\eta=1$, again, all the states between 0 and 5 MeV are
occupied with the exception of a $h_{11/2}$ state at 2.2 MeV (42\% of
occupation). This time there are several unoccupied states just above 5 MeV,
the lowest energy being at 5.4 MeV ($p_{3/2}$ state). Hence, the unperturbed
response is expected to have some structure starting from $\sim$ 10.8 MeV
with a small contribution at $\sim$ 4.4 MeV. 

For the nucleus $^{136}$Sn the situation is different: there are several 
low-lying unoccupied states. For $\eta=0.35$ the lowest energy for a
completely unoccupied state is 1.9 MeV ($p_{3/2}$ state). At 0.8 MeV one
also finds a $f_{7/2}$ state with 45\% of occupation.  In the case $\eta=1$
the lowest energy for a totally unoccupied state is 1.7 MeV ($p_{3/2}$
state) and a $f_{7/2}$ state is found at 1.6 MeV with 32\% of occupation. 
The unperturbed response is expected to start at $\sim$ 3.8 and 3.2 MeV for 
$\eta=0.35$ and 1, respectively. In the former case a small contribution at
$\sim$ 1.6 MeV is also expected.

\begin{table}[!t]
  \caption{\label{124occ} Neutron quasiparticle states with $E \le $  6 MeV 
and occupation less than 80\%. The nucleus is $^{124}$Sn. }
    \begin{ruledtabular}
      \begin{tabular}{ccccc}
 $\eta$  &  State    &  $E$ (MeV) & occ    \\
       \hline
	0.35 &  h11/2 & 1.5 & 0.42 \\
     &   f7/2 & 5.8 & 0.01     \\
\hline
  0.65 &  h11/2 & 1.7 & 0.42 \\
     &   f7/2 & 5.7 & 0.01     \\
\hline
 1  & h11/2 & 2.2 & 0.42 \\
   &  p3/2 & 5.4 & 0.003 \\
   &  f7/2 & 5.5 & 0.02 \\
   &  p1/2 & 5.6 & 0.002 \\
   &  s1/2 & 5.7 & 0.002 \\
      \end{tabular}
    \end{ruledtabular}
\end{table}

\begin{table}[!t]
  \caption{\label{136occ} Same as in Table II but for $^{136}$Sn. }
    \begin{ruledtabular}
      \begin{tabular}{ccccc}
 $\eta$  &  State    &  $E$ (MeV) & occ    \\
       \hline
	0.35 &  f7/2 & 0.8 & 0.45 \\
     &   p3/2 & 1.9 & 0.01     \\
    &    p1/2 & 2.4 & 0.006 \\
     &   f5/2 & 2.9 & 0.01 \\
    &    s1/2 & 3.3 & 0.0005 \\
    &   d5/2 &  4.0 & 0.0002 \\
    &  d3/2 &  4.1 & 0.0005 \\
    &  g9/2 & 5.6 &  0.0001 \\
    &  g7/2 & 5.6 &  0.0001 \\
\hline
  0.65  &  f7/2 & 0.9 & 0.43 \\
     &   p3/2 & 1.9 & 0.02     \\
    &    p1/2 & 2.4 & 0.008 \\
     &   f5/2 & 2.9 & 0.02 \\
    &    s1/2 & 3.2 & 0.0004 \\
    &   d5/2 &  3.9 & 0.0001 \\
    &  d3/2 &  3.9 & 0.0003 \\
    &  g9/2 & 5.4 &  0.0001 \\
    &  g7/2 & 5.5 &  0.00003 \\ 
\hline
 1  & f7/2 & 1.6 & 0.32 \\
   &  p3/2 & 1.7 & 0.02 \\
   &  p1/2 & 1.9 & 0.01 \\
   &  s1/2 & 1.9 & 0.0004 \\
   &  d5/2 & 2.6 & 0.0003 \\
   &  d3/2 & 2.6 & 0.0002 \\
   &  f5/2 & 3.0 & 0.01 \\
   &  g9/2 & 4.1 & 0.0002 \\
   & g7/2  & 4.1 & 0.0001 \\
      \end{tabular}
    \end{ruledtabular}
\end{table}

\begin{figure}
\begin{center}
\epsfig{file=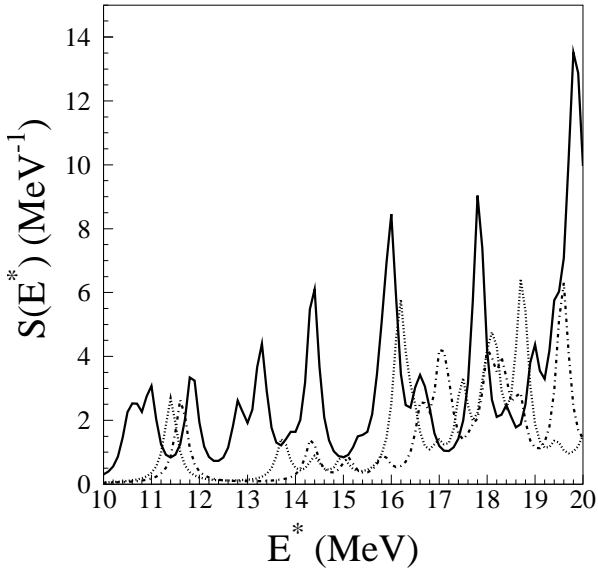,width=8.5cm}
\end{center}
\caption{Unperturbed response function for $^{124}$Sn in the two neutrons
0$^+$ addition mode. The pure surface mode is in solid line, the $\eta$=0.65
mode is in dotted line, and the $\eta$=0.35 mode in dashed-dotted lines}
\label{fig:g0124}
\end{figure}

In order to disentangle between the various pairing effects, the unperturbed
response in the two neutrons addition mode is first shown on Fig.
\ref{fig:g0124} for $^{124}$Sn. The unperturbed response is built on the HFB
single quasiparticle (QP) spectrum, for the three pairing interactions. It
should be noted that the spectrum is showed above 10 MeV, because there is
only the h$_{11/2}$ subshell which can welcome two neutrons to make a low
energy state: all the other configurations belong to the next major shell
(see Table II), explaining this high energy feature of the spectrum, as
stated above. For all the mixed pairing interaction, the unperturbed spectrum
is similar, showing that both the single quasiparticle energy and wave
functions are close to each other in that case. However, in the case of the
pure surface pairing, the spectrum is changed. The energies are shifted to
lower values, and the overall strength is increased. The lower energy shift
can be understood by more single QP states located at low energy. This can
be explained by a lower pairing gap and a different energy spectrum found in
the HFB self-consistent procedure. The larger magnitude comes from the wave
functions, and will be studied with the QPRA response. It can be already
stated that the QRPA response will also have more strength at lower energy,
due to this peculiar feature of the unperturbed spectrum for the pure
surface pairing force.

\begin{figure}
\begin{center}
\epsfig{file=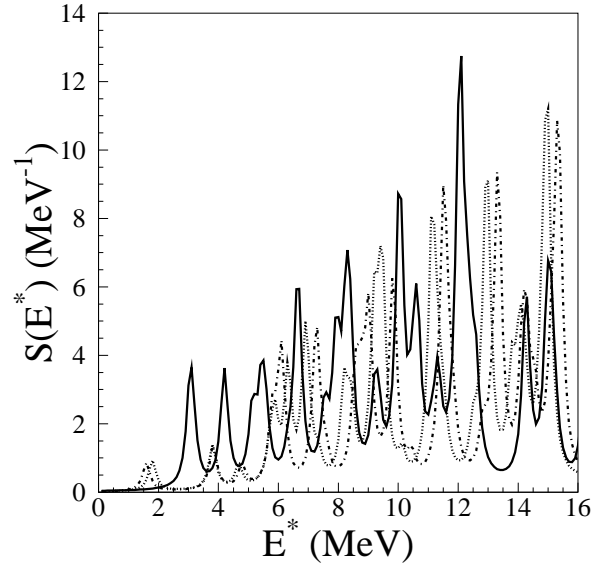,width=8.5cm}
\end{center}
\caption{Unperturbed response function for $^{136}$Sn in the two neutrons
0$^+$ addition mode. The pure surface mode is in solid line, the $\eta$=0.65
mode is in dotted line, and the $\eta$=0.35 mode in dashed-dotted lines.}
\label{fig:g0136}
\end{figure}

Fig. \ref{fig:g0136} shows the unperturbed response for the two neutron
addition mode in $^{136}$Sn. In this case, at the beginning of an open
neutron shell several low energies configurations can welcome the two
neutrons (see Table III). As in the case of $^{124}$Sn, the response
exhibits larger strength at low energy in the specific case of the pure
surface pairing interaction, compared to others pairing interaction. This is
related to the pairing field profile as shown on Fig. \ref{fig:densi}. It
should be noted that in order to clearly see the effect due to the surface
pairing, not only the first 0$^+$ state, but also the energy area of a few
MeV above should be explored since the results are different from 0 to 4 MeV
on Fig. \ref{fig:g0136}.

\subsection{Perturbed response results}

\begin{figure}
\begin{center}
\epsfig{file=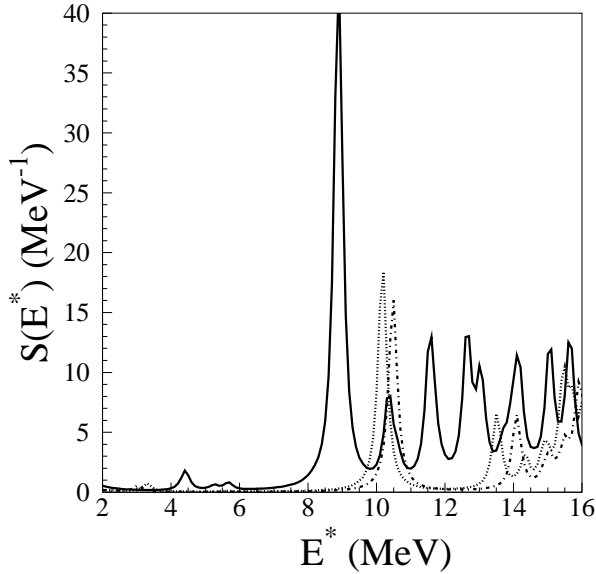,width=8.5cm}
\end{center}
\caption{QRPA response function for $^{124}$Sn in the two neutrons
0$^+$ addition mode. The pure surface mode is in solid line, the $\eta$=0.65
mode is in dotted line, and the $\eta$=0.35 mode in dashed-dotted lines.}
\label{fig:g124}
\end{figure}

Fig. \ref{fig:g124} shows the QRPA response for $^{124}$Sn, with a pure
surface and the two mixed interactions. As expected the residual interaction
plays a similar role in all the cases, gathering strength and shifting it to
lower energy. In the case of $^{124}$Sn, a peak around 9 MeV is the
strongest for the surface pairing interaction, to be compared with the one
around 10 MeV for the other interactions. Hence it is expected that the
pairing vibration transition strength should be larger in the case of a pure
surface force. However it is known that it is difficult to describe
accurately the magnitude of these transitions, especially for absolute cross
section calculations \cite{ig91}: one-step or sequential two-step process,
triton wave function, zero-range or finite-range DWBA have to be considered.
It is therefore necessary to rely on the angular distribution, calculated
from the form factor, related itself to the pairing transition density
\cite{oer01}.

The pairing transition density is defined as: 
\begin{equation}\label{eq:rhotildr} 
\kappa^\nu\left({\bf r},\sigma\right) = 
\left<0|c\left({\bf r},\bar{\sigma}\right) 
c\left({\bf r},\sigma\right)|\nu\right> 
\end{equation} 
where $c^{\dagger}\left({\bf r},\bar{\sigma}\right)$= 
$-2\sigma c^{\dagger}\left({\bf r},-\sigma\right)$ is its time reversed 
counterpart. 

\begin{figure}
\begin{center}
\epsfig{file=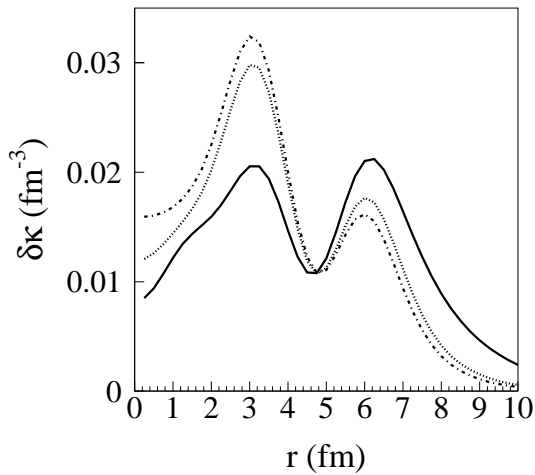,width=8.5cm}
\end{center}
\caption{Neutron transition density in the two neutrons addition mode for
$^{124}$Sn for the first peak located at 9-10 MeV. The pure surface mode 
is in solid line, the $\eta$=0.65
mode is in dotted line, and the $\eta$=0.35 mode in dashed-dotted lines.}
\label{fig:denst}
\end{figure}

It allows to calculate the form factor in the zero-range DWBA approximation.
The pairing transition densities of Fig. \ref{fig:denst} show, in the case
of $^{124}$Sn, a difference, going from surface to other modes: the
transition density decreases at the surface. However the difference is not
dramatic and may be overruled by the experimental uncertainties. The larger
strength of the 9 MeV peak in the pure surface pairing interaction is due to
a larger density at the surface.

\begin{figure}
\begin{center}
\epsfig{file=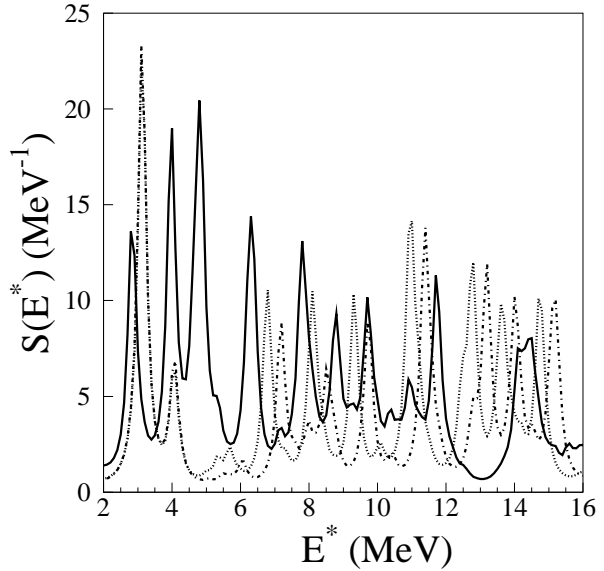,width=8.5cm}
\end{center}
\caption{QRPA response function for $^{136}$Sn in the two neutrons
0$^+$ addition mode. The pure surface mode is in solid line, the $\eta$=0.65
mode is in dotted line, and the $\eta$=0.35 mode in dashed-dotted lines.}
\label{fig:136g}
\end{figure}

For the $^{136}$Sn neutron-rich nucleus, the low energy spectrum displayed on
Fig. \ref{fig:136g} is dramatically changed from using surface to other
interactions, on a several MeV area. A three peaks structure appears in the
surface case, compared to the 2 peak structure of the other cases. The
integrated strength is also larger in the surface case.

\begin{figure}
\begin{center}
\epsfig{file=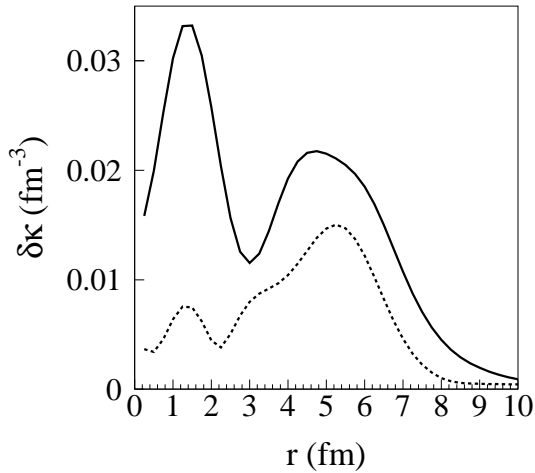,width=8.5cm}
\end{center}
\caption{Neutron transition density in the two neutrons addition mode for
$^{136}$Sn for the first two peaks of the strength, in the case of the mixed
$\eta$=0.65 interaction}
\label{fig:x65}
\end{figure}

\begin{figure}
\begin{center}
\epsfig{file=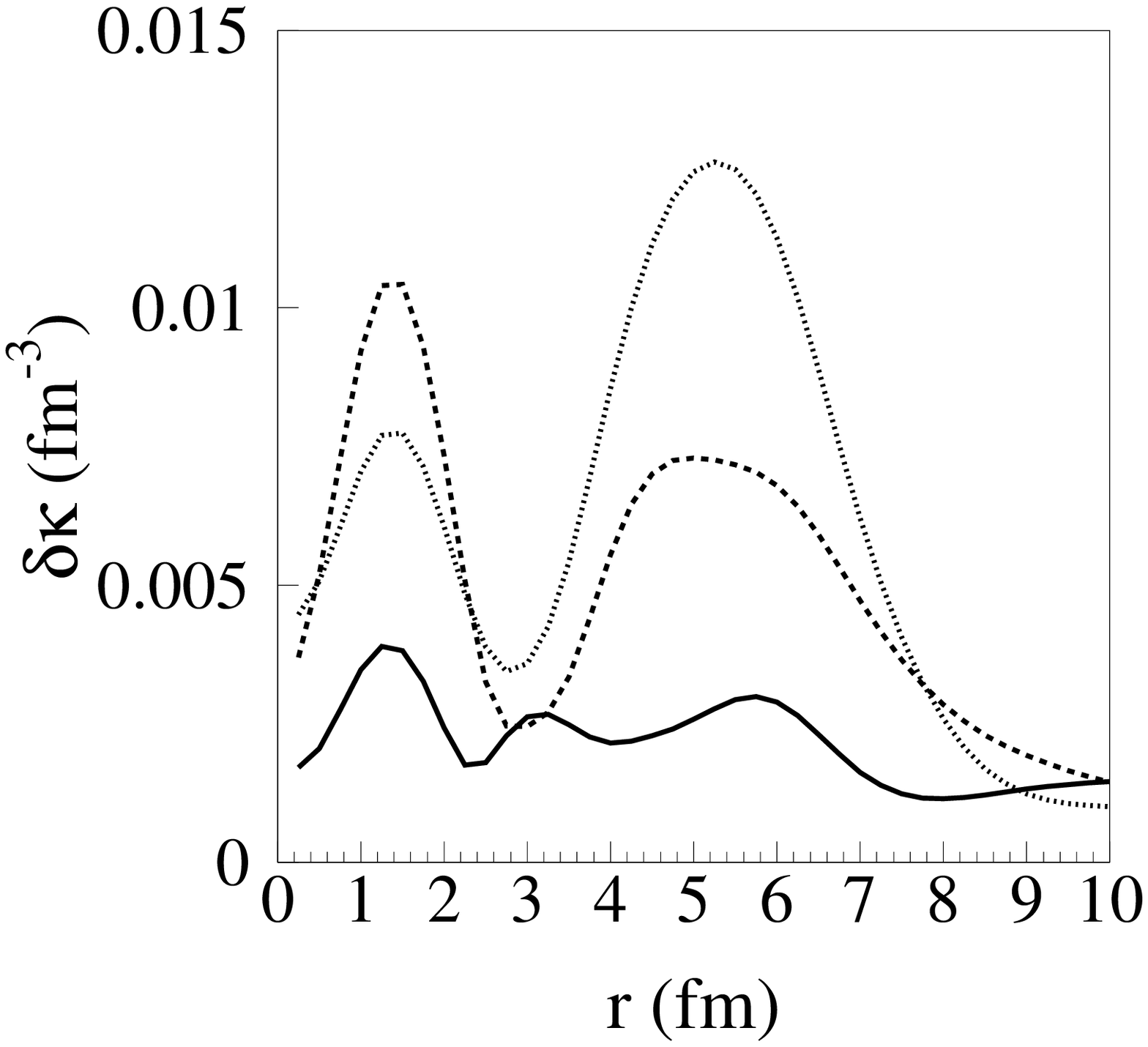,width=8.5cm}
\end{center}
\caption{Neutron transition density in the two neutrons addition mode for
$^{136}$Sn for the first 3 peaks of the strength, in the case of the pure
surface interaction}
\label{fig:x1}
\end{figure}

Fig. \ref{fig:x65} and \ref{fig:x1} show the corresponding transition
densities. They exhibit very different shapes, comparing results with the
pure surface pairing interaction and the mixed pairing interaction. Hence
$^{136}$Sn is a good test case to probe the pairing interaction through
pairing vibrations. For instance in the case of the most intense peak, the
central part is dominant in the transition density for the mixed case,
whereas the surface part of the transition density dominates in the pure
surface interaction. Hence a measurement of the angular distributions
associated with the pairing vibration strength in very neutron rich-nuclei
such as $^{136}$Sn seems more decisive to disentangle between the pairing
interactions than with $^{136}$Sn. This may be due to the larger neutron
skin in $^{136}$Sn, consisting of low density neutron-rich matter.

It has been shown in a previous work how the pairing transition densities
allow to calculate the two neutron form factor in order to predict angular
distributions for pairing vibrations \cite{kha04}. Work along these lines
should be undertaken in order to bring additional constrains on the
pairing interaction.


\section{Conclusions}

The impact of various pairing interactions on pairing vibrations predictions
has been analysed for the first time using a HFB+QRPA approach. They should
provide a good sensitivity from a pure surface interaction compared to mixed
interactions, especially in the case of very neutron-rich nuclei such as
$^{136}$Sn. Moreover nuclear matter gap calculations show that the LDA in
the pairing channel is valid in the surface of the nuclei for the
$\eta=0.35$ mixed surface/volume interaction, but not rigorously valid in
other cases. In the case of exotic nuclei, pairing vibrations are found more
sensitive to the surface/volume type of the pairing interaction, than in the
case of stable nuclei. This may be due to the larger extension of the
neutron density in very neutron-rich nuclei.

The same study using an isospin dependent pairing interaction will be
undertaken. The hope is to come one step closer to a more global pairing
interaction, using odd even mass staggering, pairing vibrations, and nuclear
matter as constraints. Experimentally, the pairing transition densities can
be tested through the form factor used to calculate the two neutrons
transfer cross section. This implies to use a adequate reaction model. Work
along these lines will be undertaken in an near future.

\noindent{\bf Acknowledgements} The authors thank D. Beaumel, M. Matsuo and
A. Vitturi for fruitful discussions


\begin{thebibliography}{*}

\bibitem{dea03} D. J. Dean and M. Hjorth-Jensen, Rev. Mod. Phys. {\bf 75}, 607 (2003)
\bibitem{gar99} E. Garrido, P. Sarriguren, E. Moya de Guerra, P. Schuck,      
{\it Phys. Rev} {\bf C60} (1999) 064312.               
\bibitem{doba}J. Dobaczewski, H. Flocard, and J. Treiner, Nucl. Phys. 
{\bf A 422}, 103 (1984)
\bibitem{gor07} S. Goriely, M. Samyn, and J. M. Pearson, {\it Phys.
Rev.} {\bf C75}, 064312 (2007)
\bibitem{pil07} N. Pillet, N. Sandulescu, P. Schuck, {\it Phys. Rev.} {\bf C76}
024310 (2007).
\bibitem{mat05} M. Matsuo, K. Mizuyama, Y. Serizawa, {\it Phys. Rev.} {\bf C71}
064326 (2005).
\bibitem{hag07a} K. Hagino and H. Sagawa, Phys. Rev. {\bf C76}, 047302 (2007)
\bibitem{mat06} M. Matsuo, {\it Phys. Rev.} {\bf C73} 044309 (2006).
\bibitem{hag07} K. Hagino, H. Sagawa, J. Carbonell, and P. Schuck
 {\it Phys. Rev. Lett.} {\bf 99}, 022506 (2007)
\bibitem{mar07} J. Margueron, H. Sagawa, K. Hagino, {\it Phys. Rev.} {\bf C76}
064316 (2007).
\bibitem{mar08} J. Margueron, H. Sagawa, K. Hagino, {\it Phys. Rev.} {\bf C77}
054309 (2008).
\bibitem{bro73} R. A. Broglia, O. Hansen and C. Riedel, {\it Advances in
Nuclear Physics, NY Plenum} {\bf 6}(1973) 287.
\bibitem{oer01} W. von Oertzen, A. Vitturi, {\it Rep. Prog. Phys.} {\bf
64}(2001) 1247.
\bibitem{kha04} E. Khan, N. Sandulescu, Nguyen Van Giai, M. Grasso, {\it Phys.
Rev.} {\bf C69} 014314 (2004).
\bibitem{ave08} B. Avez, C. Simenel, Ph. Chomaz, {\it Phys. Rev.} {\bf C78}
044318 (2008).
\bibitem{cha98} E. Chabanat, P. Bonche, P. Haensel, J. Meyer, R. Schaeffer, 
{\it Nucl. Phys.} {\bf A635} (1998) 231.
\bibitem{kha02} E. Khan, N. Sandulescu, M. Grasso, N. V. Giai, 
{\it Phys. Rev.} {\bf C66} (2002) 024309.  
\bibitem{mat09b} M. Matsuo, {\it Proc. of COMEX 3 Conference} (2009) 
\bibitem{mat09} K. Mizuyama, M. Matsuo and Y. Serizawa 
{\it Phys. Rev.} {\bf C79} (2009) 024313.  
\bibitem{ig91} M. Igarashi, K. Kubo and K. Yagi, {\it Phys. Rep.} 
{\bf199}(1991) 1.
\end{thebibliography}
\end{document}